\documentclass{article}

\usepackage{arxiv}
\usepackage{xcolor}

\usepackage[utf8]{inputenc} 
\usepackage[T1]{fontenc}    
\usepackage{hyperref}       
\usepackage{url}            
\usepackage{booktabs}       
\usepackage{amsfonts}       
\usepackage{nicefrac}       
\usepackage{microtype}      
\usepackage{lipsum}
\usepackage{graphicx}
\usepackage{titlesec}
\usepackage{amsmath}
\usepackage{float}
\usepackage[euler]{textgreek}
\graphicspath{ {./images/} }
\usepackage[noadjust]{cite}

\title{Deep Multi-Branch CNN Architecture for Early
Alzheimer’s Detection from Brain MRIs

}

\author{
 Paul K. Mandal* \\
  Department of Computer Science\\
  University of Texas at Austin\\
  Austin, TX 78712 USA \\
  \texttt{mandal(at)utexas.edu} \\
  \\
  *Corresponding author\\
   \And
 Rakeshkumar Mahto \\
  Department of Electrical and Computer Engineering\\
  California State University Fullerton\\
  Fullerton, CA 92831 USA \\
  \texttt{ramahto(at)fullerton.edu} \\
    \And
  {\footnotesize \textsuperscript{}for the Alzheimer’s Disease Neuroimaging Initiative}\thanks{Data used in preparation of this article were obtained from the Alzheimer’s Disease Neuroimaging Initiative
(ADNI) database (adni.loni.usc.edu). As such, the investigators within the ADNI contributed to the design
and implementation of ADNI and/or provided data but did not participate in analysis or writing of this report.
A complete listing of ADNI investigators can be found at:
http://adni.loni.usc.edu/wp-content/uploads/how\_to\_apply/ADNI\_Acknowledgement\_List.pdf}
}

\begin{document}
\maketitle
\begin{abstract}
Alzheimer's disease (AD) is a neuro-degenerative disease that can cause dementia and result severe reduction in brain function inhibiting simple tasks especially if no preventative care is taken. Over 1 in 9 Americans suffer from AD induced dementia and unpaid care for people with AD related dementia is valued at \$271.6 billion. Hence, various approaches have been developed for early AD diagnosis to prevent its further progression. In this paper, we first review other approaches that could be used for early detection of AD. We then give an overview of our dataset that was from the Alzheimer’s Disease Neuroimaging Initiative (ADNI) and propose a deep Convolutional Neural Network (CNN) architecture consisting of 7,866,819 parameters. This model has three different convolutional branches with each having a different length. Each branch is comprised of different kernel sizes. This model can predict whether a patient is non-demented, mild-demented, or moderately demented with a 99.05\% three class accuracy.
\end{abstract}

\keywords{
Alzheimer's, Brain Imaging, CNN, Convolution, Convolutional Neural Network, Deep Learning, Disease Detection, Neural Network, Machine Learning, Medical Diagnosis
}

\section{Introduction}
\label{sec:introduction}
Alzheimer's Disease (AD) is a common disease that affects 1 in 9 (10.7\%) Americans over 65. 6.5 million Americans aged 65 or over have been diagnosed with AD dementia. An estimated 16 billion unpaid man-hours of care were given to people with dementia from AD in 2021 which has an estimated value of \$271.6 billion\cite{b1}. Approximately 12\% to 18\% of people over 60 are living with mild cognitive impairment (MCI)\cite{b2}. MCI causes subtle changes in memory and thinking. Although often associated with the normal aging process, MCI is not apart of typical aging. Moreover, 10\%-15\% of individuals with MCI develop full dementia each year\cite{b3}. Therefore, AD must be diagnosed at an early stage to prevent it from progressing further. For this purpose, Machine Learning (ML) and Deep Learning (DL) can play an invaluable role since they have been extensively used in various other medical applications for diagnosing and detecting various abnormalities and diseases\cite{R1, R2, R3}.  

A diverse range of approaches has been employed in the field of early Alzheimer's disease (AD) detection, encompassing the analysis of speech patterns and inflections, neuropsychometric tests, olfactory tests, eye testing, gait testing, as well as the utilization of neural networks on various diagnostic modalities such as MRIs, electroencephalograms (EEG), and magnetoencephalographs (MEG) \cite{b5}. Recently, there has been a surge in the popularity of machine learning and deep learning techniques for early AD diagnosis, with a predominant focus on applying these methods to MRI images, MEG, EEG, and other relevant physiological parameters\cite{R17}.

\subsubsection{Neural networks on MRIs}

Over the years, various neural network techniques have been used to predict Alzheimer’s. In \cite{R5}, a convolution neural network (CNN) was applied to accurately predict mild cognitive impairment to AD. The overall accuracy reported in \cite{R5} was around 86.1\%. A similar technique of CNN was applied to a dataset consisting of MRI images of 156 AD and 156 normal patients \cite{R6}. The dataset in this study consisted of AD patients and age/gender-matched normal individuals \cite{R6}. The proposed technique in \cite{R6} achieved an accuracy of 94\%. A mix of 2D CNN and recurrent neural networks (RNN) on MRI images were reported to achieve an accuracy of 96.88\% \cite{R7}. The proposed technique applied an RNN after applying a 2D CNN to recognize the connection between 2D image slices \cite{R7}. The study also presented a technique of transfer learning from 2D images to 3D CNNs. 
One of the best-performing models that didn’t rely on MRIs was a neural network trained to analyze speech patterns. One of their models reported a 97.18\% accuracy \cite{b6}. However, there are two main issues with this approach. First and foremost is that it is clear from looking at the audio waves that the subjects who have AD are well past the MCI/Mild Demented stage making it non-viable for an early detection stage. The second is that the study only had 50 non-demented subjects and 20 demented subjects. Each non-demented subject had 12 hours of audio and each demented subject had 8 hours of audio. These clips were divided into 600 different clips of 60 second audio. However, the paper does not state whether they divided the training and validation sets by patient or not. If that is the case, there is a possibility that the neural network is learning how to classify whether the subject has AD based on the patient’s voice, rather than extracting useful information.

\subsubsection{Neural networks on magnetoencephalographs (MEG)}
Compared to MRI images, a non-invasive diagnostic technique called Magnetoencephalography (MEG) is utilized for measuring brain activity. Based on brain activity, the proposed method estimates the magnetic field generated by the slow ionic current flow through cells. This research shows that MEG activity can provide excellent sensitivity for early diagnosis of DP \cite{R9}. A combination of the MEG recording and MRI scans are utilized in \cite{R10}, which resulted in an accuracy of 89\%. A similar technique for diagnosing AD was presented in \cite{R11}. However, the accuracy of the classification technique was 77\%. Various other machine learning (ML) driven techniques for diagnosing AD using MEGs are summarized in \cite{R12}. However, none of the techniques were able to achieve an accuracy greater than than 90\%.

\subsubsection{Neural networks on electroencephalograms (EEG)}
Another more promising study in AD diagnosis has been done with EEGs. Electrophysiological imaging techniques such as EEGs are widely accepted as reliable indicators for the diagnosis of AD. With the aid of neural networks, it has become possible to use EEG data to accurately determine whether a patient has Alzheimer's disease. A novel neural network, I-Fast, was able to predict whether subjects had AD with 92\% accuracy \cite{R13}. The dataset used in this study consisted of 115 mild cognitive impairment and 180 AD patients. This is significant as it implies that EEGs can be used as a viable alternative for the diagnosis of AD, given the cost effectiveness of the technique. Similarly, a novel technique was presented in \cite{R14} that uses a finite response filter (FIR) in a double time domain to extract features from an EEG recording dataset consisting of MCI, AD, and healthy controls (HC). Later, Binary Classification (BC) achieved an accuracy of 97\%, 95\%, and 83\% between HC vs. AD, HC vs. MCI, and MCI vs. AD, respectively.

\subsubsection{Blood Plasma}
Another approach was to test for a panel of 18 different proteins from blood samples. This approach was able to achieve an 89\% accuracy\cite{b7}. This is probably the most promising of the methods described above since it is much easier and less costly to run blood tests. Although we do concede that there are benefits to the techniques outlined above due to the limited availability of MRIs, none of the approaches enumerated above were able to exceed our 99\% accuracy achieved from our approach.


\begin{figure}[H]
\center
\includegraphics[width=0.4\columnwidth]{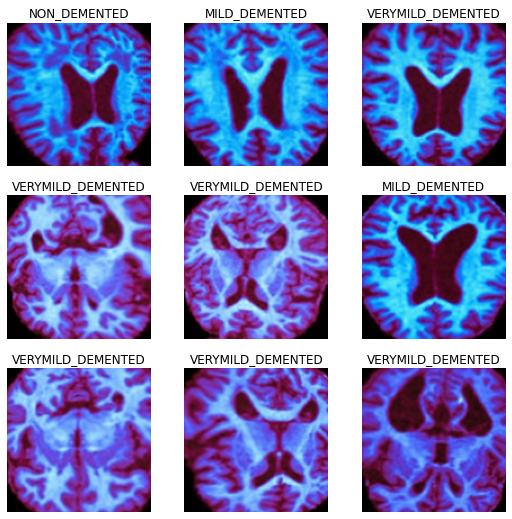}
\center
\caption{A sample of 9 preprocessed images from the ADNI dataset.}\label{fig1}
\end{figure}

\subsection{Description of Alzheimer's MRI Datasets}
Our dataset consists of 6338 magnetic resonance imaging (MRI) images that were imaged from the Alzheimer’s Disease Neuroimaging Initiative (ADNI)\cite{b4} and were curated and preprocessed on Kaggle\cite{b22}. Our preprocessed dataset came formatted in 100x100 pixel images. Our dataset consists of 3202 images of non-demented patients, 2242 images of very mild demented patients, and 892 images of mildly demented patients as we opted to not use the 64 moderate demented images due to the low sample size. Figure \ref{fig1} shows 9 images from the ADNI dataset.

\begin{figure}[H]
\center
\includegraphics[width=0.5\columnwidth]{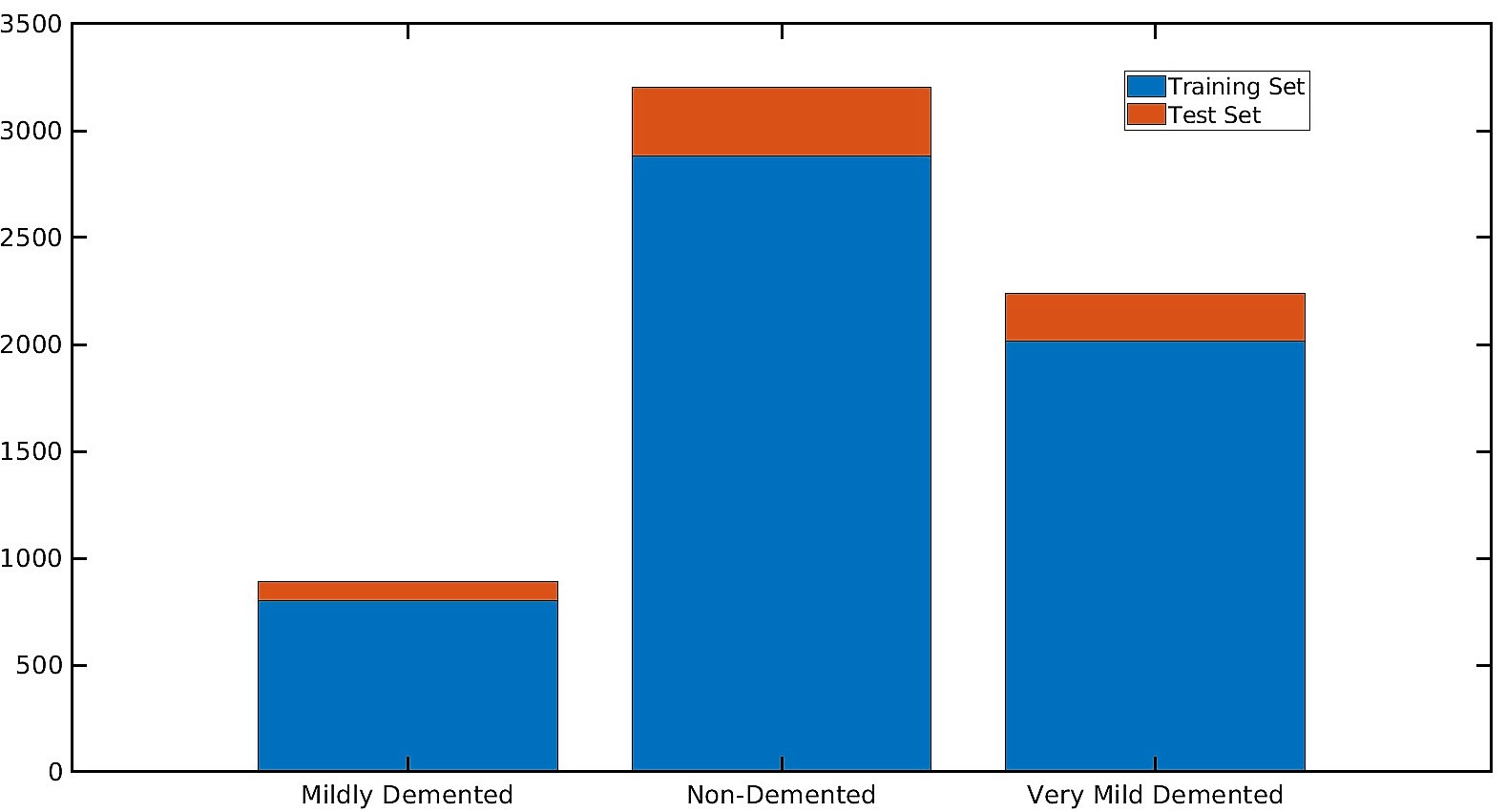}
\center
\caption{Distribution of Training and Test images from the ADNI dataset. }\label{fig1b}
\end{figure}

We divided our training and test set into 5701 training images (2881 non-demented, 2017 very mild demented, and 802 mild demented) and 637 test images (321 non-demented, 225 very mild demented, and 90 mild demented) as shown in Figure \ref{fig1b}. We used stratified random sub-sampling to ensure that the training and test sets had the same ratio of non-demented, very mild demented, and mild demented images. 

\section{Background}
Significant gains have been made in image recognition and object detection through the use of deep learning. These advances have been applied to medical imaging such as diabetic retinapathy \cite{b9}. For the purposes of this paper, familiarity of the subsequent concepts is necessary.

\begin{figure}[H]
\center
\includegraphics[width=0.6\columnwidth]{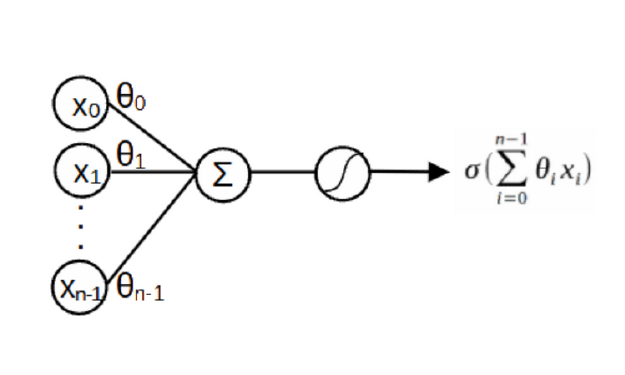}
\caption{A simple neuron classifier with a sigmoid activation function.}\label{fig2}
\end{figure}

\subsection{The Neuron as a binary classifier}
A neuron in a neural network is a mathematical model loosely based on how a neuron works in the brain\cite{b10}. A neuron in a brain works by receiving electric signals to it's dendridic tree. There is a positive or negative synaptic weight which determines whether the signal will be sent to the axon. If the sell fires, the signal is then sent through the cell's axon to other dendritic trees.

An artificial neuron works in a similar manner. A neuron takes a series of inputs and performs an inner product with the weight vector. This value is then put through a sigmoid activation function. This activation function then produces an output between zero and one which represents the percent confidence that the neuron has that the input belongs to a certain category as shown in Figure \ref{fig2} \cite{b11}.

We define the sigmoid activation function to be,
\begin{equation}\sigma=\frac{1}{1+e^{-z}}.\label{sigmoid}\end{equation}

We can train an optimal weight vector, \texttheta, by minimizing the cost function \begin{equation}J(\theta)= - \frac{1}{m} \sum_{i=1}^{m} Cost(\theta^{T}x^{i},y^{i})\label{train}\end{equation}

Where our cost function, $Cost(\theta^{T}x,y)$ is defined as,

\begin{equation}
Cost(\theta^{T}x,y)=
    \begin{cases}
        -log(\sigma(\theta^{T}x)) & \text{if } y = 1 \\
        -log(\sigma(1 - \theta^{T}x)) & \text{if } y = 0
    \end{cases}
\end{equation}

\subsection{Feed Forward Neural Networks and the Multi Layer Perceptron}

\begin{figure}[H]
\center
\includegraphics[width=.5\columnwidth]{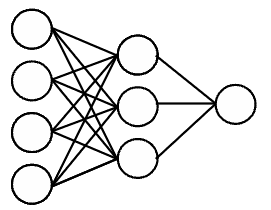}
\center
\caption{A three layer feed forward network.}\label{fig3}
\end{figure}

Neural networks can extract useful features by representing the data. These intermediate features can then be used for classification, regression, or decoding by subsequent layers \cite{b13}. A feed forward neural network is the most simple deep learning architecture. A feed forward neural network is when a layer of neurons forms a fully connected bipartite graph with the next layer. A typical neural network will have an input layer, and output layer, and a certain amount of hidden layers as shown in Figure \ref{fig3}.

There are severe limitations of feed forward neural networks when applied to images. First and foremost, a regular feed forward neural network is not invariant under translation. Thus, other approaches are required for better results.

\subsection{Convolutional Neural Networks}

We define the convolution operation, $*$, between two discrete functions $f[m,n]$ and $g[m,n]$ to be:

\begin{equation}(f * g)[m,n] = 
    \sum_{i = -\infty}^{\infty} \sum_{j = -\infty}^{\infty} f[m,n] g[m-i,n-j]
\end{equation}

Convolution is a useful operation in signal processing. Convolution can apply filters to images that can result in different effects, such as sharpening or smoothing \cite{b14}. In a neural network, our convolutional layers have a few parameters: the number of filters, the kernel size which determines the size of each filter, and a stride size which determines how many pixels to "skip" before applying the filter to the next block of pixels.

Assuming a stride size of 1, for a given $N \times M$ image, we apply an $i \times j$ filter. We divide our image into $(N - i + 1) \times (M - j + 1)$ subimages and perform an inner product between the filter and the subimage. The result of the convolution is a filtered image with $(N - i) \times (M - j)$ pixels.

A convolutional network learns optimal filters by training filter parameters by minimizing our loss \cite{b15}. These trained filters can learn useful filters that can help subsequent layers classify our data. This is applicable to many sorts of problems such as facial recognition \cite{b16}. 

\subsection{Pooling}

Pooling is a method of reducing the dimension for the outputs of convolutional layers. This has two main advantages. First, it helps prevent convolutional networks from overfitting. Secondly, if applied properly, it can decrease the computational expense of a CNN. For this reason, pooling is often used in neural networks for mobile systems.\cite{b17}.

There are two main kinds of pooling, max-pooling and average-pooling. Max-pooling will take the largest value in a kernel\cite{b18}. Average pooling instead takes the average of all of the outputs within a kernel \cite{b19}. Average-pooling will produce a smoother output than max-pooling will.

\section{Methods}
The neural network architecture in the study was implemented utilizing the widely-used Keras libraries \cite{b20} which are seamlessly integrated into TensorFlow \cite{b21}. Through extensive experimentation, it was discovered that a multiple-branch approach, incorporating varied kernels, demonstrated the best performance on the specific dataset under investigation. The multiple-branch model achieved an accuracy rate of approximately 99.05\%, although this is a marginal improvement over its single-branch counterpart which had achieved accuracy of 98.7\%.

Regarding the hardware configuration, the system used for the study was equipped with an NVIDIA GeForce RTX 2080S graphics card boasting 8GB of VRAM. The computational power was complemented by an Intel Xeon processor, specifically the Intel Xeon W-10855M, featuring a total of 24 logical cores and 12 physical cores. To support efficient processing and memory management, the system had 64 GB of main memory.

In terms of the training process, the neural network underwent approximately 100 epochs to optimize its parameters. The training duration typically lasted for approximately 3-4 hours when applied to the ADNI Dataset. The Adam optimizer was employed with a learning rate of 0.001, facilitating efficient convergence and exploration of the dataset.

\subsection{Layers}
In this subsection, an overview of the layers used in the neural network architecture is provided, along with the parameters they take.

\subsubsection{Input}
The neural network architecture accepted a 100x100x3 input tensor. In this context, the image dimensions were 100x100, composed of pixels, and each pixel had 3 color values ranging from 0 to 255. To ensure consistency, the data was normalized by dividing these color values by 255, resulting in floating-point values.

\subsubsection{Convolutional Block}
A block consisting of three layers was defined and utilized throughout the architecture. The first layer in this block was a 2-D convolutional layer. Here, the number of filters to be generated by the convolutional layer and the specified kernel size were determined. Following the convolutional layer, an average pooling layer was applied. Lastly, the block concluded with a regularization layer to enhance generalization and prevent overfitting.

\subsubsection{Dropout Layer}
Dropout is a technique used in deep learning to prevent overfitting. This is done by randomly selecting a specified percentage of neurons to not train on during each epoch.

\subsubsection{Dense Layer}
The dense layer, known as a fully connected layer, played a crucial role in the neural network architecture employed in the study. As discussed in the background section, this layer consisted of neurons that were interconnected with all the neurons from the preceding layer. This connectivity allowed for information propagation and integration across the network, enabling the network to learn complex patterns and relationships within the data.

\subsection{Model Architecture}

The network architecture employed in this study comprised three distinct branches. The first branch utilized convolutional layers with a 3x3 kernel size and an average pooling size of 2x2. Within this branch, five convolutional blocks were implemented, generating 32, 64, 128, 256, and 512 filters, respectively. The resulting output was then flattened, and a dropout rate of 50\% was applied.

Moving on to the second branch, it employed convolutional layers with a 5x5 kernel size and an average pooling size of 3x3. Three convolutional blocks were included in this branch, generating 128, 256, and 512 filters. Similar to the previous branch, the output was flattened and subjected to a dropout rate of 50\%.

The third branch utilized convolutional layers with a 7x7 kernel size and an average pooling size of 5x5. This branch encompassed two convolutional blocks, generating 128 and 256 filters. Consistent with the previous branches, the output was flattened, and a dropout rate of 50\% was applied.

The outputs of the three branches were concatenated, creating a unified layer. This concatenated layer was then fed into a feed-forward neural network with 256 outputs. Subsequently, a dropout rate of 50\% was applied, followed by an additional feed-forward layer with 128 outputs. Once again, a dropout rate of 50\% was implemented. Finally, the architecture culminated in a final layer comprising three output neurons, trained using the softmax function. Each neuron corresponded to a specific class label: non-demented, very mild demented, or mild demented.

\begin{figure}
\center
\includegraphics[width=.8\columnwidth]{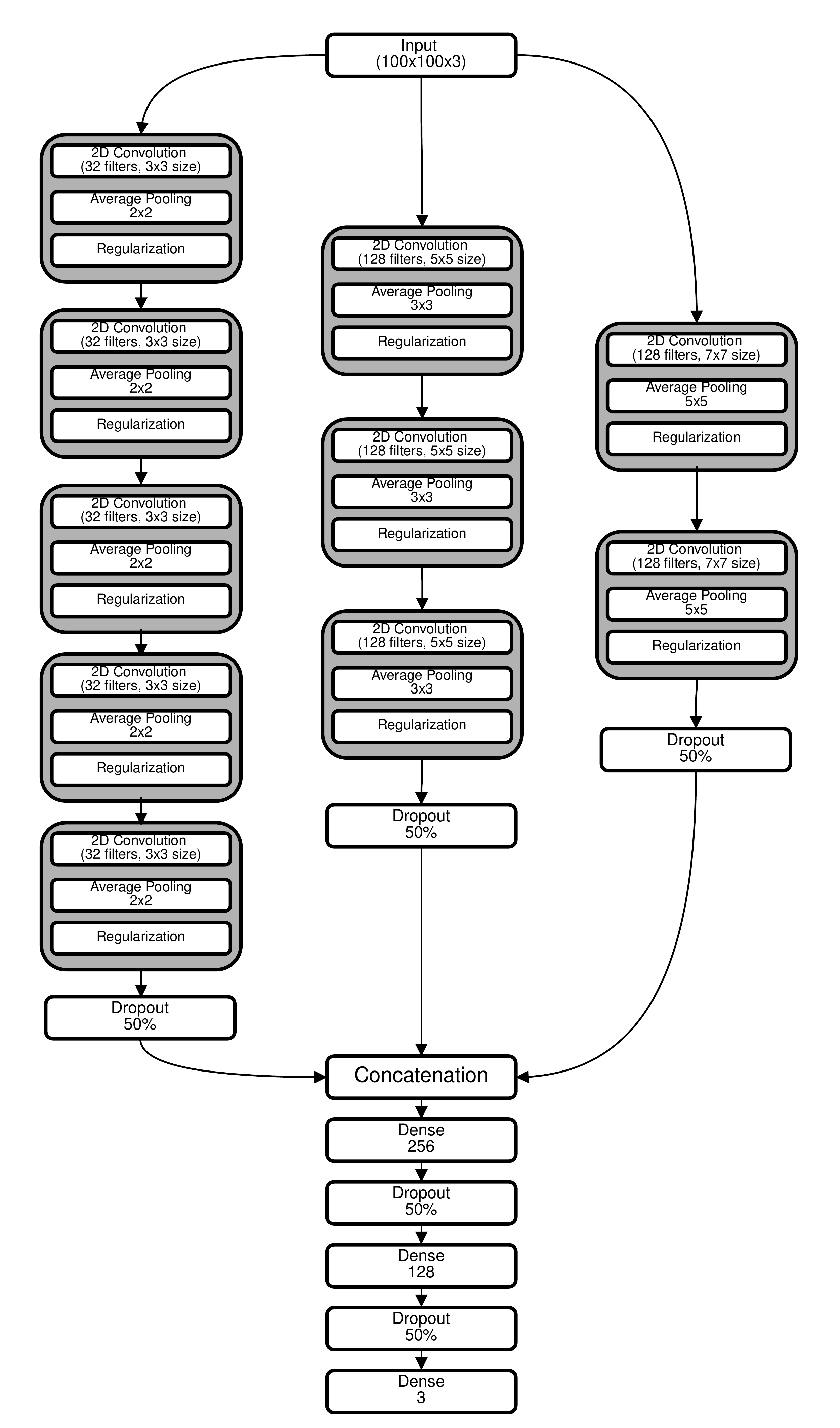}
\center
\caption{Our proposed network architecture.}\label{fig4}
\end{figure}

\section{Experimental Results and Discussion}

\begin{figure}

\includegraphics[width=.8\columnwidth]{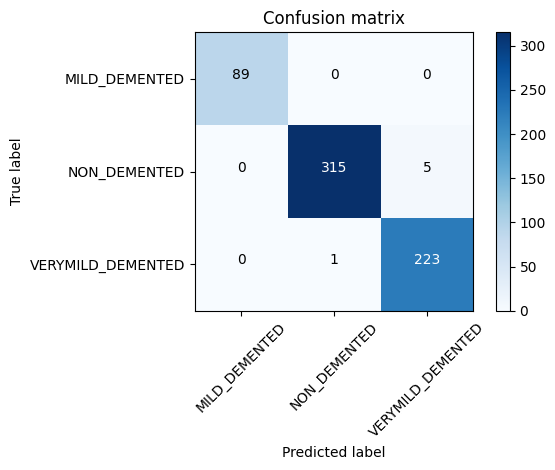}
\center
\caption{A confusion matrix showing the results of our network.}\label{fig5}
\end{figure}

Our neural network achieved an accuracy of 99.05\%.\footnote{Code and weights are available here in Modelv11.ipynb: https://github.com/PaulKMandal/Alzheimer-Detection} Of our 637 test images, only 6 were misclassified with all of these images being between the Non-Demented and Very Mild Demented classes. We found that our neural network did not overfit and witnessed that validation accuracy would cease to improve after about 100 epochs. Thus, after finding our model that best fit our validation data, we trained on both the training and validation sets and then ran the test set through our neural network.

Despite our exceptionally high accuracy, there are many things that we could do to make this more applicable to a clinical setting. The first would be to create a new method to preproccess our Alzheimer's data rather than using the possibly outdated data uploaded to Kaggle. During our preprocessing, we separated the layers of the MRI into images and trained on each individual image. Further insight could be derived by comparing how our classifier performed on different layers of the same MRI or alternatively comparing different MRIs on the same patient taken at different periods. 

Although designing a classifier that trained on the entire MRI instead of each layer might yield a lower accuracy because of the lower sample size, it might prove to be more useful in the clinical setting. Alternatively, one could train a neural expert system by inputting each layer from the MRI into a branch of a neural network and then concatenating the outputs of those branches and training it on a subsequent neural network. Finally, implementing a Fully Convolutional Network to highlight areas of the brain that are indicators could potentially have some use.

\begin{figure}[H]
  \begin{center}
 \includegraphics[width=.8\columnwidth]{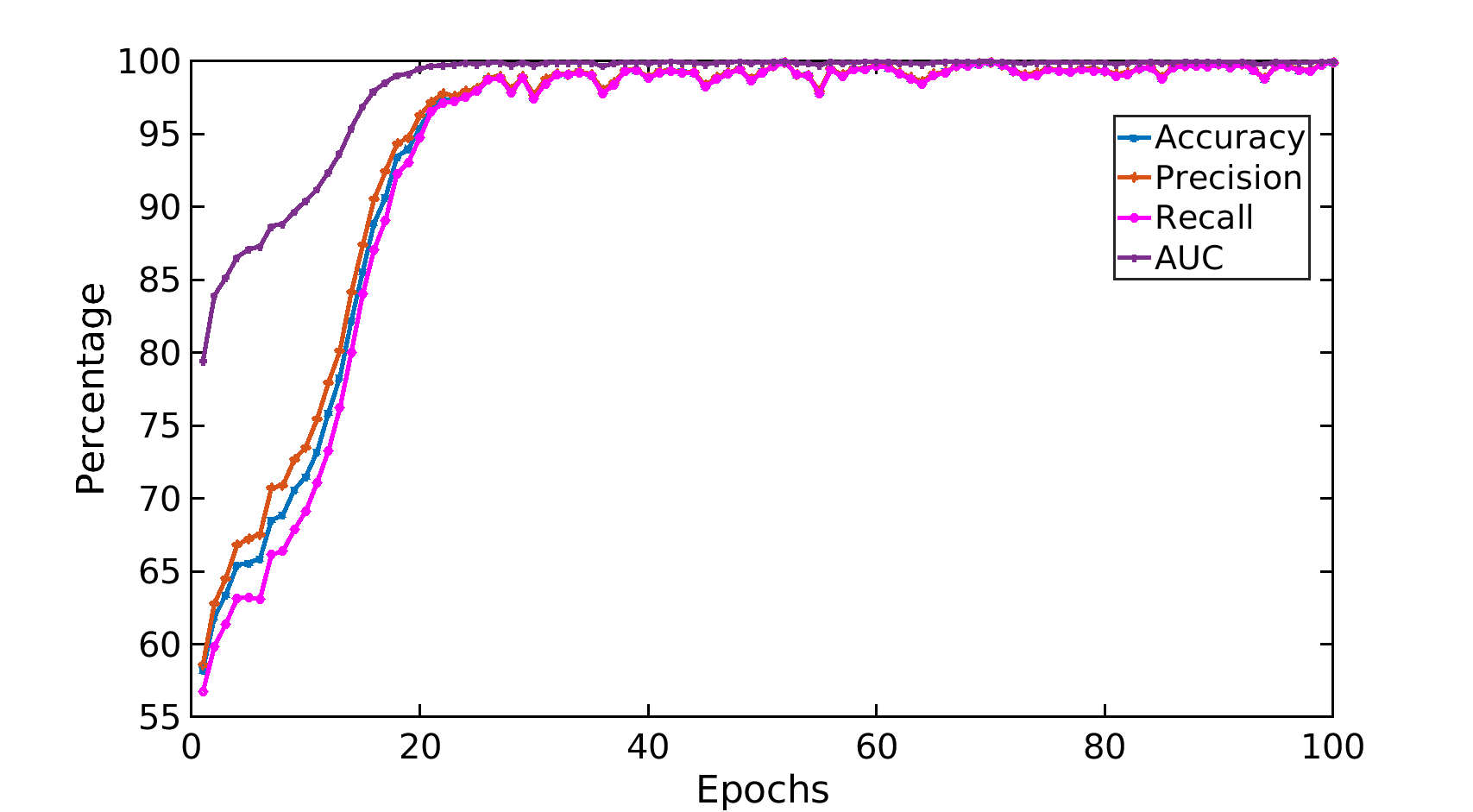}\\
  \caption{Improvement in accuracy, precision, recall, and AUC with an increase in Epochs}\label{Fig7}
  \end{center}
\end{figure}

It would also be useful to augment our data from The Open Access Series of Imaging Studies (OASIS)\cite{b23}. Attempting to prepossess our data in a way that is as agnostic to the MRI machine as possible could potentially provide much more use in the medical community.

\section{Conclusion}

An early diagnosis of Alzheimer's disease is a critical challenge facing society today. The disease not only results in a severe reduction in brain function, but also puts a significant burden on caregivers and the healthcare system. The results of this study demonstrate the potential of using deep learning techniques, specifically Convolutional Neural Networks, to improve the early diagnosis of AD. The proposed architecture has achieved high accuracy of  99.05\%  in predicting the progression of the disease and is a step forward in developing tools for 

However, the current approach has its limitations and several directions that this research could take to make it more practical and useful in a clinical setting. For example, it would be beneficial to improve pre-processing methods across a more general dataset of MRI images to ensure that the data used more closely fits the patient population. Additionally, further exploration of the classifier’s performance on different layers of the MRI could provide valuable insights into the disease progression. One promising avenue for future work is to implement a neural expert system that inputs each layer of the MRI into a separate branch of a network and concatenates the outputs for further analysis. Another possibility is to use a Fully Convolutional Network to highlight areas of the brain that are indicative of AD, which could provide important information to clinicians in making a diagnosis.

Furthermore, more complex models could be created that combine our neural network architecture with the other methods enumerated in the Previous Research section in order to achieve even higher accuracy classifiers. One field of inquiry that could be useful would be to build a federated expert system where a heterogenous federated machine learning model is designed so that each "expert" trains on a different federate, but such a project is well outside the scope of this paper.

\section*{Acknowledgment}

The authors of this paper would like to acknowledge Aloke K. Mandal, M.D., Ph.D. for his help on explaining the differences between various medical definitions such as MCI and Mild Dementia.

Data collection and sharing for this project was funded by the Alzheimer's Dise ase
Neuroimaging Initiative (ADNI) (National Institutes of Health Grant U01 AG024904) and
DOD ADNI (Department of Defense award number W81XWH-12-2-0012). ADNI is funded
by the National Institute on Aging, the National Institute of Biomedical Imaging and
Bioengineering, and through generous contributions from the following: AbbVie, Alzheimer’s
Association; Alzheimer’s Drug Discovery Foundation; Araclon Biotech; BioClinica, Inc.;
Biogen; Bristol-Myers Squibb Company; CereSpir, Inc.; Cogstate; Eisai Inc.; Elan
Pharmaceuticals, Inc.; Eli Lilly and Company; EuroImmun; F. Hoffmann-La Roche Ltd and
its affiliated company Genentech, Inc.; Fujirebio; GE Healthcare; IXICO Ltd.; Janssen
Alzheimer Immunotherapy Research \& Development, LLC.; Johnson \& Johnson
Pharmaceutical Research \& Development LLC.; Lumosity; Lundbeck; Merck \& Co., Inc.;
Meso Scale Diagnostics, LLC.; NeuroRx Research; Neurotrack Technologies; Novartis
Pharmaceuticals Corporation; Pfizer Inc.; Piramal Imaging; Servier; Takeda Pharmaceutical
Company; and Transition Therapeutics. The Canadian Institutes of Health Research is
providing funds to support ADNI clinical sites in Canada. Private sector contributions are
facilitated by the Foundation for the National Institutes of Health ( www.fnih.org). The grantee
organization is the Northern California Institute for Research and Education, and the study is
coordinated by the Alzheimer’s Therapeutic Research Institute at the University of Southern
California. ADNI data are disseminated by the Laboratory for Neuro Imaging at the
University of Southern California.

\end{document}